\documentstyle[12pt]{article}

\newcommand{\be}{\begin{equation}}
\newcommand{\ee}{\end{equation}}
\newcommand{\eel}[1]{\label{#1}\end{equation}}
\newcommand{\bea}{\begin{eqnarray}}
\newcommand{\eea}{\end{eqnarray}}
\newcommand{\eeal}[1]{\label{#1}\end{eqnarray}}
\newcommand{\baq}{\begin{equation}\begin{array}{rcl}}
\newcommand{\eaq}{\end{array}\end{equation}}
\newcommand{\eaql}[1]{\end{array}\label{#1}\end{equation}}
\newcommand{\beac}{\begin{equation}\begin{array}{rcl}}
\newcommand{\eeacn}[1]{\end{array}\label{#1}\end{equation}}
\newcommand{\ba}{\begin{array}}
\newcommand{\ea}{\end{array}}
\newcommand{\non}{\nonumber \\}

\renewcommand{\a}{\alpha}

\newcommand{\e}{\varepsilon}

\newcommand{\al}{{\alpha^{'}}}
\newcommand{\beq}{\begin{eqnarray}}
\newcommand{\eeq}{\end{eqnarray}}
\newcommand{\nn}{\nonumber}

\newcommand{\journal}[4]{{\rm #1~}{#2}\,(19#3)\,#4}

\newcommand{\np}{\journal {Nucl. Phys.}}
\newcommand{\pl}{\journal {Phys. Lett.}}

\begin{document}
\begin{titlepage}

\begin{flushright}

\vspace{2cm}

TAUP-2466-97\\
hep-th/9711010\\
\end{flushright}
\vspace{2cm}

\begin{center}
{\bf\LARGE  More on Probing Branes with Branes}\footnote{
Work supported in part by the US-Israel Binational Science Foundation,
by GIF - the German-Israeli Foundation for Scientific Research, and by
the Israel Science Foundation.
}\\
\vspace{1.5cm}
{\bf A. Brandhuber, N. Itzhaki, }\\
{\bf J. Sonnenschein and S. Yankielowicz}\\
\vspace{.7cm}
{\em
Raymond and Beverly Sackler Faculty of Exact
 Sciences\\School of Physics and Astronomy\\Tel Aviv University,
  Ramat Aviv, 69978, Israel\\
e-mail: andreasb, sanny, cobi, shimonya@post.tau.ac.il}
\end{center}
\vskip 0.61 cm

\begin{abstract}
We generalize the Gibbons-Wiltshire solution of four dimensional
Kaluza-Klein black holes in order to describe Type IIA
solutions of bound  states of D6 and D0-branes.
We probe the solutions with a D6-brane and a D0-brane.
We also probe a system of D2+D0-branes and of  
a D2-brane bound to a F1-string with a D2-brane.
A precise agreement between the SYM and the SUGRA calculations  is
found for the static force as well as for the $v^2$ force in all cases.
\end{abstract}

\end{titlepage}
\baselineskip 18pt
\section{Introduction}

About a year ago it was conjectured that M-theory in the infinite
momentum frame is described by a supersymmetric matrix model
\cite{mat}.
The advantage  of  this approach is that   it  provides 
a regularization scheme  for  the M-theory 
 at short distances via  a SYM  description.
The success  of such a regularization scheme
 requires, among other things, that 
at large distances and low energies, where eleven dimensional
SUGRA is a good approximation to M theory,
the matrix model interactions  yield  SUGRA interactions. 
This was the main motivation behind the intensive study  in the last
year of  the correspondence between   the SUGRA and SYM descriptions of  
the long distance interactions of D-branes.

On the SUGRA side the (bosonic) action of a test Dp-brane in the background of
a source of Dp-branes is \footnote{We work in units where $\al =1$.}
\be
 \label{2}
S=-
\frac{1}{g(2\pi)^p}
\int d^{p+1}x  e^{-\phi}
\sqrt{\mbox{det} h_{\mu\nu}}+ A_{01...p}
\ee
where $h_{\mu\nu}$ is the induced metric, $A_{01...p}$ is a p-form. 
The action is treated classically and at large distances (compared to
the relevant length scale of the source) it can be expanded in powers
of $1/r$. On the SYM side at low energies the Born-Infeld action of $Q+1$
Dp-branes can be approximated by 10D $N=1$ SYM with $U(Q+1)$ gauge
symmetry reduced to $p+1$ dimensions
\be\label{11}
 S_0=\frac{1}{g(2\pi)^p}\int d^{p+1}x \frac{1}{4}
Tr[F_{MN}F^{MN}]+\mbox{Fer.},
\ee
where $M, N$ are ten dimensional indices. In terms of this field
theory the presence of a probe Dp-brane at distance $r$ away from a
source of $Q$ Dp-branes is equivalent to an expectation value to one of the
scalars along the flat directions. This expectation value breaks the
$U(Q+1)$ to  $U(Q)\times U(1)$. 

At the one loop order the effective SYM action was calculated for a
general gauge field background \cite{102,mal}
\be\label{3}
S_1=\frac{c_{7-p}}{8(2\pi)^pr^{7-p}}\int d^{p+1}x 
\left( TrF^4-\frac{1}{4}Tr(F^2)^2\right),
\ee
where $F_{MN}$ is the background field,
$c_q=(4\pi)^{q/2-1}\Gamma(q/2)$ and $r$ is the distance between the
Dp-brane probe and the source Dp-branes. 

A precise agreement between eq.(\ref{3}) and the leading
term  (in $1/r$) of eq.(\ref{2}) was found in many cases 
\cite{mat,mal,kps,aha,gil2,gil,200,201,202,203,204,pie,205,kra}.
For the D0 case in the infinite momentum frame 
it was shown that the field theory two loop calculation does not
contribute to the leading term interaction \cite{bec} but  produces
precisely the next to leading term in $1/r$ obtained from the SUGRA
approach \cite{poll}. Moreover, non-perturbative effects on the brane
were related, in the case of D2-branes, to scattering with momentum
transfer along the $x_{10}$ direction in \cite{pol1} (see also
\cite{matt,suu}).

The structure of eq.(\ref{3}) is similar to the structure
of the sub-leading term in the expansion of the Born-Infeld action.
This led to the conjecture \cite{40,41} that the effective action
generated by the SYM at any order for any Dp-brane will give rise to
the corresponding term in the expansion of the Born-Infeld action.
Under this assumption the correspondence between SUGRA
and SYM was proven to any order in several cases \cite{40,41}.

In this note we examine  some more examples supporting
the correspondence between SYM and SUGRA.
In sec. 2 we probe a bound state of Dp-branes and D(p-2)-branes with a
Dp-brane. Most of the results of this section are well known
\cite{aha,gil2,40}. We believe that we perform the calculation in a 
somewhat different and hopefully interesting way.
In sec. 3 we probe a bound state of a D2-brane and a F1-string
with a D2-brane. We use the M-theory description of a D2-brane and
a F1-string as unwrapped and wrapped M2-branes respectively to  
unify different ten dimensional forces.
We use T-duality to relate results for different D-brane dimensions.
In sec. 4 we generalize the solutions of Gibbons and Wiltshire of four
dimensional black holes \cite{gib} in order to describe the most general
type IIA solution which contains a bound state of D6-branes and
D0-branes. We probe this solution with  a D6-brane and compare to the  
SYM result. Applying electric-magnetic duality we relate our result 
to the result of \cite{kra}. 

In all the cases we have considered a precise agreement between SYM  
and SUGRA is obtained for the static force as well as for the $v^2$ force.
We conclude with a discussion on the possible significance of our
result for the case of D6-brane.

\section{Probing  Dp+D(p-2) branes with a Dp-brane}

The system of a bound state of a Dp-brane and a D(p-2)-brane and a probe
Dp-brane is not supersymmetric. Therefore, 
a static force as well as a force proportional to $v^2$ between
the bound state and the probe Dp-brane are expected.
We start with probing a D2-D0 system with a D2-brane and compare the
SYM result to the SUGRA result. Then we apply  T-duality along some of
the transverse directions and show that also the T-dual systems
yield the same SYM and SUGRA results.

\subsection{The SYM calculation}

The effect of a D0 brane being immersed in a D2-brane is to induce a
background magnetic field, $B=F_{12}$, on the D2 world-volume. From
eq.(\ref{3}) we see that the resulting static potential is 
\be\label{e}
 U_{stat}=\frac{3n_2 V_2 B^4}{16 r^5},
\ee
where $V_2$ is the area of the D2.
To find the leading velocity dependent potential which is proportional
 to $v^2$ one can use the expression found in \cite{mal}
\be\label{aa}
U_{v^2}=\frac{c_{7-p}g v^2 E}{2r^{7-p}},
\ee
where $E$ is the total energy on the D-brane, $E=\int d^2x T_{00}$.
In our case $E=V_2B^2/8g\pi ^2$ thus
\be\label{d}
U_{v^2}=\frac{3n_2V_2 B^2 v^2}{8 r^5}.
\ee
The $v^4$ potential does not depend on the background magnetic field
and it is obtained from eq.(\ref{3}) by taking on the probe
$F_{0i}=\partial_0x^i=v_i$
\be\label{67}
U_{v^4}=\frac{3n_2V_2  v^4}{16 r^5}.
\ee 

\subsection{ The SUGRA calculation}

We perform the calculation in  eleven dimensional SUGRA 
using the notation of \cite{pol1}.
The eleven dimensional SUGRA solution of a M2-brane with $x_{10}$ 
compactified on a circle of radius $R_{10}$ is \cite{duff}
\be\label{26}
ds^2=f^{-2/3}(-dx_0^2+dx_1^2+dx_2^2)+f^{1/3}(dx_3^2+...+dx_{10}^2),
\ee
where
\be\label{27}
f=1+r_0^6\sum_{n=-\infty}^{\infty}\frac{n_2}{(r^2+(2\pi R_{10}n)^2)^3}.
\ee
The sum is due to the images of the source membrane on the
compactified direction $x_{10}\sim x_{10}+2\pi R_{10}n$.
At large distances ($r\gg R_{10}$) we get
\be\label{f1}
f=1+\frac{3n_2 r_0^6}{16R_{10}r^5}.
\ee
Adding D0-branes to the D2-brane is the same as boosting the solution
in the $x_{10}$ direction \cite{rus}. Then we want to  probe the
boosted solution with a M2-brane moving in the 
transverse directions $x_3, ..., x_9$ with a constant velocity.
Instead we  keep the M2 source static and give the probe a velocity
along the $x_{10}$ direction. 

The action of the probe in this background is
\be\label{69} 
S=-
\int d^{3}x  \tau_2
\sqrt{\mbox{det} h_{\mu\nu}}+ \mu_2H, 
\ee
where  the induced metric  $h_{\mu\nu}$  is given by
\be
 h_{\mu\nu}=g_{\mu\nu}+\partial_{\mu}x^i\partial_{\nu}x^{j}g_{ij},
\ee
Expanding in $v$ we find that the leading term of the probe's action
in this background is 
\be\label{c}
S=\frac{3n_2V_2\tau_2 r_o^6}{128r^5R_{10}}(v^2+v_{10}^2)^2\int dx_0,
\ee
where $v^2=\sum_{i=3}^{9}v_i^2$.
Since $\tau _2 r_{0}^6=8R_{10}$ \cite{pol1} and $B=v_{10}$
\cite{tow,lu,yyy} we find a  precise agreement  with the SYM results,
eqs.(\ref{e}, \ref{d}, \ref{67}).

\subsection{T-duality}

Next we wish to show the correspondence between the
SYM and the SUGRA approaches for a bound state of Dp and D(p-2)
branes probed by D-p branen for general $p$.
We use the result of sections 2.1 and 2.2 for the $p=2$ case of 
D2-branes and D0-branes and apply T-duality.
We show explicitly that applying T-duality in the SYM description yields
the same result as in the SUGRA description.

Let us start with  the SYM approach.
The description of a bound state of a Dp-brane and a 
D(p-2)-brane is very similar to the description in the case of D2+D0.
Suppose that the world volume of the Dp-brane is along 
$x_0, x_1, ..., x_p$ and that the world volume of the D(p-2)-brane is
along $x_0, x_3, x_4, ..., x_p$. 

Then the presence of D(p-2)-branes on the Dp-brane is
translated, in the SYM theory, to a constant field background 
$F_{12}$ on the D-p brane world volume.
The calculation for Dp-D(p-2) is  a straight-forward
generalization of the D2-D0 calculation.
For instance from eq.(\ref{3}) it is clear that 
the ratio between the result for D3-D1 and the
result for D2-D0 is
\be\label{12}
\frac{S_{31}}{S_{20}}=\frac{c_4V_3r}{2\pi c_5 V_2}=\frac{2R_3 r}{3\pi}.
\ee

Let us turn now to SUGRA calculation. The D3-D1 action can be obtained
from  the D2-D0 action by T-duality.
In order to do so we need to compactify the $x_3$ direction, $x_3\sim
x_3+2\pi R_3$.
Then we should take into account also the images of the source  along
the $x_3$ direction.
This means that the harmonic function, $f(r)$,  for the compactified case
contains also the images of the source.
Therefore, the ratio between  the action
for   compactified $x_3$, $S_{20 com}$,
and  the action for  uncompactified $x_3$, $S_{20}$, is
\be\label{cv}
\frac{S_{20 com}}{S_{20}}=\sum_{m=-\infty}^{\infty}
\frac{r^5}{(r^2+(2\pi m
  R_3)^2)^{5/2}}.
\ee
At large distances the sum on the right hand side of eq.(\ref{cv}) 
can be replaced by an integral and we obtain
\be
\frac{S_{20 com}}{S_{20}}=\frac{2r}{3\pi R_3}.
\ee
Since T-duality takes $R_3\rightarrow 1/R_3$ we get a precise
agreement with eq.(\ref{12}).

\section{Probing D2+F1 with a D2-brane}

\subsection{The SUGRA calculation}

Again we perform the computation in eleven dimensions.
The origin of a  F1-string and a 2D-brane in M-theory is a wrapped and
an unwrapped M2 brane respectively.
Therefore, a bound state of a F1-string along, say, $x_1$ (smeared in
the $x_2$ direction) and a D2-brane can be described in M theory as a
M2-brane at an angle $\alpha_{10}$ relative to the probe which
contains no  F1-string. The angle   $\a_{10} $ is given by
\be
\tan \a_{10}=\frac{\partial x_{10}}{\partial x_2}=\sigma_{F1},
\ee
where $\sigma_{F1}$ can be thought of as the density of the 
fundamental strings on the D2.
To describe the action of the probe in this background we can
rotate the source (eqs.(\ref{26}, \ref{f1}))
in the $x_2$, $x_{10}$ plane while keeping the probe
fixed
or we can keep the source fixed and rotate the probe.
We follow the second approach. 

In that case we can use the SUGRA solution eq.(\ref{26},\ref{f1}) and  
eq.(\ref{69}) to find for $\a, \a_{10}\ll 1$
\be\label{30a}
U=\frac{3n_2V_2(\a^2 +\a_{10}^2)^2}{16r^5},
\ee
where $\a^2=\sum_{i=3}^{9}\a_i^2$ and $\tan \a_i=\frac{\partial
  x_{i}}{\partial x_2}$. 
Note that just like the velocity in the previous section the angles in
this section break super-symmetry and hence a static force is generated.

\subsection{The SYM calculation}

The relation between $\a$ and $F_{\mu\nu}$ on the D2-brane is given by 
\cite{tow,lu,yyy}
\be
E_1=F_{01}=\frac{\partial x_{10}}{\partial x_2}=\tan \a_{10}.
\ee
A trivial generalization (taking $F_{2i}=\a_i$, $i=3,...9$) of the result
 of subsection 2.1 gives for $\a\ll 1$
\beq
&&U_0=\frac{3n_2 V_2 E_1^4}{16 r^5},\non
&&U_{\a^2}=\frac{3n_2V_2 E_1^2 \a^2}{8 r^5},\\
&&U_{\a^4}=\frac{3n_2 V_2 \a^4}{16 r^5},\nonumber
\eeq
which is in a precise agreement with eq.(\ref{30a}).

Note, that T-duality along $x_1$  transforms the source into a bound  
state
of D1-brane and F1-string with momentum along the $x_1$ direction and
the probe  into  a  D1 brane.
Then the $\a^4$ force between the source and the probe, in the D2+F1
case, becomes a  $v^4$ force in the D1-brane case \cite{rus}\footnote{We thank
  A.A. Tseytlin for pointing this out to us.}.

\section{Probing D6+D0-branes with a D6-brane}

\subsection{ The SUGRA calculation}

There are two possible approaches to compute the Type IIA solution for
a bound state of  D6-branes and D0-branes.
The first one is to start with the six-brane solution in type IIA 
\cite{st} (not necessarily extremal) and to lift it to eleven
dimension, then boost it along the $x_{10}$ direction (which means
that we add D0's in the type IIA language) and  reduce it back to ten
dimensions. The second approach is to generalize the Gibbons-Wiltshire (GW) 
solution \cite{gib} of the  four dimensional black hole with electric and
magnetic charges which solve the five dimensional Einstein-Hilbert
vacuum equation.
The relation to a D6+D0 bound state solution is anticipated since in
type IIA a D0-brane has an electric charge and a D6-brane has magnetic charge.

We shall follow the second approach.
Our starting point is  the eleven dimensional
Einstein-Hilbert action
\be
S=\frac{1}{(2\pi)^8g^3}\int d^{11}x\sqrt{-g_{11}}R_{11}.
\ee
The dimensional reduction
\be
ds_{11}^2=e^{4\phi /3}(dx_{11}+A_{\mu}dx^{\mu})+e^{-2\phi /3}
ds^2_{10},
\ee
leads to the type IIA action (in the string frame)
\be
S=\frac{1}{(2\pi)^7g^2}\int
d^{10}\sqrt{-g_{10}}\left( e^{-2\phi} 
(R_{10}+4(\nabla\phi)^2 ) -\frac14 F^2\right ). 
\ee
We have kept only fields which are relevant to the solutions
 considered below.

The spherically symmetric time independent 
solutions are parameterized by the  mass (M), the electric charge
(Q) and the magnetic charge (P).
$P$ is proportional to $n_6$ while $Q$ is proportional to $n_0$.
The exact relation will be discussed soon.
The dilaton charge $\Sigma$ is related to $M, Q, P$ by
\be\label{20} 
\frac{8}{3}\Sigma=\frac{Q^2}{\Sigma +\sqrt{3}M}
+\frac{P^2}{\Sigma -\sqrt{3}M}.
\ee
The Type IIA generalization of the GW solution is 
\beq
&& e^{4\phi / 3}=\frac{B}{A},\non
&& A_{\mu}dx^{\mu}=\frac{Q}{B}(r-\Sigma)dt+
P\cos \theta d\phi,\\ 
&& g_{\mu\nu}dx^{\mu}dx^{\nu}=-F/\sqrt{AB}dt^2+\sqrt{\frac{B}{A}}
(dx_1^2+...+dx_6^2)+
\non &&\sqrt{AB}/F dr^2
+\sqrt{AB}(d\theta^2+\sin ^2\theta d\phi ^2),\nn
\eeq 
where
\beq 
&&F=(r-r_+)(r-r_-),\non
&&A=(r-r_{A+})(r-r_{A-}),\\ 
&&B=(r-r_{B+})(r-r_{B-}),\nn
\eeq
and 
\beq
&& r_\pm=M\pm\sqrt{M^2+\Sigma^2-P^2/4 -Q^2/4}, \non
&&r_{A\pm}=\Sigma/\sqrt{3}\pm\sqrt{\frac{P^2\Sigma /2}
{\Sigma-\sqrt{3}M}},
\\ 
&&r_{B\pm}=-\Sigma/\sqrt{3}\pm\sqrt{\frac{Q^2\Sigma /2}{\Sigma+
\sqrt{3}M}}\nn .
\eeq
Under the electric-magnetic duality (D6$\leftrightarrow$D0) 
the solutions are transformed in the following  way
\beq\label{b1} 
&&Q\leftrightarrow P, \non
&&\Sigma \leftrightarrow -\Sigma,\\
&& M \leftrightarrow M.\nonumber
\eeq
The self-dual ($Q=P$) solution was obtained in \cite{she,jjj} and was
probed with D-branes in \cite{pie}. 

For the extremal cases (which we focus on in this note) it is
convenient to make the coordinate change $r\rightarrow r-M$.
Then $F=r^2$ and the metric has the simple form
\be\label{fds}
ds^2=-f_1(r)dt^2+f_2(r)dx_idx_i+f_1^{-1}(r)(dr^2+r^2d\Omega),
\ee
where 
\be
f_1(r)=r^2/\sqrt{AB},\;\; f_2(r)=\sqrt{B/A}.
\ee
To determine the exact relation between $P$ and $n_6$ and between
$Q$ and $n_0$ we compare these solutions to the well known p-brane
solution of type IIA \cite{st}.
We start with the pure magnetic extremal solution.
Namely, we take 
\be\label{21}
P=4M ,\;\; Q=0,\;\; \Sigma =-\sqrt3 M.
\ee
This solves eq.(\ref{20}) and the extremality condition and leads to
\beq\label{c1}
&&
e^{-2\phi}=f^{3/2},\non
&&
A_{\mu}dx^{\mu}=4M\cos\theta d\phi,\\
&&
ds_{10}^2=f^{-1/2}(-dt^2+dx_1^2+...+dx_6^2)+f^{1/2}(dr^2+r^2d\Omega),
\nn
\eeq
where
\be
f=1+\frac{4M}{r}.
\ee
This is obviously a D6-brane carrying magnetic charge.
Since in type IIA $f=1+gn_6/2r$ we find for the  Kaluza-Klein monopoles 
\be\label{25}
M=\frac{gn_6}{8},
\ee
which is indeed the expected relation between the mass and the
 radius of the compactified direction  \cite{pol,gro,sor} since $R_{10}=g$.
Using eq.(\ref{21}) we obtain
\be\label{c5}
P=\frac{gn_6}{2}.
\ee
Now we turn to the relation between $n_0$ and $Q$.
We consider the pure electric extremal  solution which is dual to
 eq.(\ref{21}), 
\be\label{c3}
P=0 ,\;\; Q=4M,\;\; \Sigma =\sqrt3 M,
\ee
and leads to an electric field associated with D0-branes smeared over $V_6$
\beq\label{c1a}
&&
e^{-2\phi}=f^{-3/2},\non
&&
A_{\mu}dx^{\mu}=\frac{4M}{r}dt,\\
&&
ds_{10}^2=f^{-1/2}(-dt^2+dx_1^2+...+dx_6^2)+f^{1/2}(dr^2+r^2d\Omega),
\nn
\eeq
where again
\be
f=1+\frac{4M}{r}.
\ee
In type IIA,  for D0-branes smeared along $V_6$, we have
$f=1+gn_0(2\pi)^6/2V_6r$ (see for example
section 2.4 in \cite{mall}).
Thus, we find for the pure electric extremal solution 
\be\label{yu}
M=\frac{gn_0 (2\pi)^6}{8V_6}.
\ee
To see that this is the expected result for the Kaluza-Klein momentum
we recall that we work in units where $l_s=1$. 
In units  where  $l_P^{11}=1$, $l_s=g^{-1/3}$ and $V_6\rightarrow
 g^{6/3}V_6$.
Then eq.(\ref{yu}) yields the correct KK momentum dependence,  $\sim  
1/R_{10}$.
Now we can use eq.(\ref{c3}) to get
\be\label{c4}
Q=\frac{gn_0 (2\pi)^6}{2V_6}.
\ee
It should be emphasized that although  eqs.(\ref{c5},
\ref{c4}) were derived by considering special cases they
hold in general simply because $P$ is proportional to $n_6$ and $Q$
is proportional to $n_0$.

Now we are in a position to describe the solution with a large number
of D6-branes and a small number of D0-branes ($P\gg Q$).
To do so we need to move away from eq.(\ref{21}) along eq.(\ref{20})
and the extremality condition.
We fix $M$ and take 
\be
\Sigma=M(-\sqrt3+\e ),\;\;\e\ll 1.
\ee
To leading order in $\e$ the solutions for $Q(\e )$ and $P(\e )$ are
\be\label{29}
Q^2=\frac{2\e^3}{3\sqrt3}M^2,\;\;P=(4-\sqrt3\e+\frac{\e^2}{8})M.
\ee
Using these relations and eqs.(\ref{2}, \ref{fds}) we find that
at large distances   the
 static part of the action is
\be\label{22}
S=\frac{-V_6}{(2\pi)^6gr}\left( M(-4+\sqrt3 \e)+P\right) \int
dt=\frac{-V_6}{8(2\pi)^6gr}M\e^2\int dt,
\ee
and  the term in the action which is proportional to $v^2$ is 
\be\label{ad}
S_{v^2}=\frac{\sqrt3 M V_6 \e}{2(2\pi)^6r}v^2\int dt.
\ee

\subsection{The SYM calculation}

The gauge theory which lives on the world-volume of the six brane is a
$6+1$ dimensional theory.
The SYM theory in $6+1$  is non-renormalizable and therefore 
ill defined at short
distances.
Yet we managed to use the one loop effective action of eq.(\ref{3})
to derive meaningful results for the $r$-dependent part of the force
between the source and the probe.
We recall that we are considering the case in which the $U(Q+1)$ gauge
symmetry is broken down to $U(Q)\times U(1)$ with $r$ the expectation
value of the appropriate adjoint scalar.
The fields with one index in $U(Q)$ and the other in $U(1)$ are
becoming massive with $m\sim r$.
Thus, at the one loop order the $r$-dependent contribution to the
effective action arise through the masses of the states of the open
string which are integrated out.
The remnants of these integrations are determinants
of the form $\mbox{det} (\partial_{\mu}\partial^{\mu} +m^2_i)$ (for
the bosonic fields) where $m_i$ depends on $r$ and the background.
The $r$ dependent part of these determinants is always well defined.

We use the construction of \cite{tay} which describes a bound state of
four D6 and D0-branes with no D2 and D4-brane. The background is
\beq\label{ab}
&&F_{12}=F_0 \mbox{diag}(1, 1, -1, -1),\non
&&F_{34}=F_0\mbox{diag}(1, -1, -1, 1),\\
&&F_{56}=F_0\mbox{diag}(1, -1, 1, -1),\nonumber
\eeq
One can check that this background carries no D2-branes 
($\int \mbox{Tr} F\wedge F=0$) no D4-branes ($\int \mbox{Tr} F=0$)
but does carry D0-branes ($\int \mbox{Tr} F\wedge F\wedge F\neq 0$).
Plugging this  background in eq.(\ref{3}) we get
\be\label{d1}
S=\frac{(6-6^2/4)V_6n_6}{16(2\pi)^6r}F_0^4\int dt
=\frac{-3V_6n_6}{16(2\pi)^6r}F_0^4\int dt,
\ee
where we have embedded the $U(4)$ solution of 
 \cite{tay} in $U(n_6)$ ($n_6>4$). 
Since we work in units where $\al =1$ 
\be
n_0=\frac{1}{6(2\pi)^6}\int d^6x \mbox{Tr}F\wedge F\wedge F
\ee
eqs.(\ref{c5}, \ref{c4}), imply that
\be\label{wha}
F_0^3=\frac{Q}{P}.
\ee
 Using eqs.(\ref{25}, \ref{29}, \ref{wha}) to leading order in $\e$,
a precise agreement with eq.(\ref{22}) is found. 

To find the term which is proportional to $v^2$ we  use 
eqs.(\ref{aa}, \ref{ab})
\be\label{qw}
S_{v^2}=\frac{3F_0^2n_6V_6}{8r(2\pi)^6r}v^2\int dt.
\ee
which again agrees with the  SUGRA calculation eq.(\ref{ad}).

Now we would like to use the electric-magnetic duality, 
eq.(\ref{b1}), to make contact with refs. \cite{gil,kra}.
Under electric-magnetic duality our configuration becomes a 
D0-brane probe in the background of D0+D6-branes.
The action for this probe should be identical to the action of a
D0+D6-brane 
probe moving in the background of a D0-brane which was considered in
\cite{kra}. Eq.(\ref{b1}) takes $Q\leftrightarrow P$ and hence
$F_0\leftrightarrow 1/F_0$,  and $n_6\leftrightarrow n_0 (2\pi)^6/V_6$.
It also takes the D6-brane probe to a D0-brane probe and therefore
interchanges their respective tensions $V_6/g(2\pi)^6\leftrightarrow 1/g$.
Therefore, it takes both eq.(\ref{22}) and eq.(\ref{d1}) to
\be
S=\frac{3(2\pi)^6 n_0}{16V_6 F_0^4r}\int dt=\frac{3n_6}{16 F_0r }\int dt,
\ee
which is in agreement with the result of \cite{gil,kra} (where $n_6=4$).
Acting with electric-magnetic duality on the term which is
proportional to $v^2$, eq.(\ref{qw}), yields
\be
S=\frac{3n_6F_0v^2}{8r }\int dt,
\ee
which is in agreement with the result of \cite{gil}.

\section{Conclusions}

The agreement between SUGRA and SYM has been verified by now in a
multitude of examples. It would be interesting to find the deeper
string theoretic reason for this observation.
In ref. \cite{40} it is pointed out that although by adding an
F-background SUSY is broken  the underlying supersymmetry may still
lead to the same $F^4$ term at small and large
distances in string theory. Small distances are controlled by the
(weakly coupled) SYM theory while the long distance by the SUGRA theory.
It would certainly be important to further explore this agreement.

One might argue that the computation above supports the
conjecture that there
is a ``dual'' description (in the sense of $g\rightarrow 1/g$) for 
SYM in $6+1$ dimensions which decouples from the bulk, perhaps
 along the lines of \cite{egkr,lms,bk,hg}.
The  argument  goes as follows:
Introducing explicitly the $\a^{'}$ dependence in eq.(\ref{11}) we
get
\be\label{opo}
g^2_{YM}=(2\pi)^pg\a^{{'(p-3)}/2}.
\ee
The low energy limit of superstring theory, which is relevant for our
computation of SUGRA is obtained by $\al\rightarrow 0$.\footnote{There
  are also $\al$ corrections to the Born-Infeld action.
However, these corrections involve derivatives of $F_{\mu\nu}$ and
hence they vanish for the  backgrounds considered in the present paper.} 
For the SYM computation we should also keep the gauge 
coupling constant finite.
For $p>3$ this implies that $g\rightarrow\infty$ which in turn implies
that the theory does not decouple from gravity.
If this is indeed the case we would expect also the closed string
sector associated with gravity to be relevant.
Hence, we would not expect, in these cases, the dynamics of the
Dp-brane to be described by just the Born-Infeld (or SYM) theory
associated with the open string sector.
Therefore, there seems to be no reason for the agreement
between the results of SUGRA and SYM theories for $p>3$.
However such agreements have been found in refs. \cite{pie,gil,kra}
and in the present note.
The way to explain these agreements bearing in mind the discussion above
is to claim that  for $p>3$ there is a ``dual'' description for the
dynamics of the Dp-brane which decouples from the bulk and regulates
the SYM theory at short distances.
For  $p=4,5$ such theories have been suggested \cite{brs,sei}
in the closely related
investigation of M(atrix) theory compactified on $T^4$ and $T^5$.
For $p=6$ it is not clear whether a ``dual'' description which decouples
from the bulk exists \cite{sen,sei2}.
It is, therefore, natural to argue that the fact that we are
accumulating examples which demonstrate that SYM
gives results which agree with the SUGRA result also in the $p=6$
case, support the existence of a ``dual'' description of the D6-brane
which decouple from gravity. This theory will also be the theory
describing M(atrix) theory on $T^6$.

\vspace{1cm}

We would like to thank A.A. Tseytlin for helpful comments.


\begin{thebibliography}{99}

\bibitem{mat} T. Banks, W. Fischler, S.H. Shenker and L. Susskind,
  Phys. Rev. D55 (1997) 5112.
\bibitem{102} R.R. Metsaev and A.A. Tseytlin, Nucl. Phys. B298 (1988) 109.
\bibitem{mal} J. M. Maldacena, hep-th/9709099.
\bibitem{kps} M.R. Douglas, D. Kabat, P. Pouliot and S.H. Shenker,
  Nucl. Phys. B485 (1997) 85.
\bibitem{aha} O. Aharony and M. Berkooz, Nucl. Phys. B491 (1997) 184.
\bibitem{gil2} G. Lifschytz and S.D. Mathur, Nucl. Phys. B507 (1997) 621,
 hep-th/9612087. 
\bibitem{gil} G. Lifschytz, hep-th/961223.
\bibitem{200} V. Balasubramanian and F. Larsen, Nucl. Phys. B506 (1997) 61, 
hep-th/9703039.
\bibitem{201} D. Berenstein and R. Corrado, Phys. Lett. B406 (1997) 37, 
 hep-th/9702108.
\bibitem{202} M. Douglas, J. Polchinski and A. Strominger, hep-th/9703031.
\bibitem{203} I. Chepelev and A.A. Tseytlin, Phys. Rev. D56 (1997) 3672,
 hep-th/9704127.
\bibitem{204} I. Chepelev and A.A. Tseytlin, hep-th/9705120.
\bibitem{pie} J. M. Pierre, Phys. Rev. D56 (1997) 6710, hep-th/9707102.
\bibitem{205} R. Gopakumar and S. Ramgoolam, hep-th/9708022.
\bibitem{kra} E. Keski-Vakkuri and P. Kraus, hep-th/9706196.
\bibitem{bec} K. Becker and M. Becker, Nucl. Phys. B506 (1997) 48,
 hep-th/9705091.
\bibitem{poll} K. Becker, M. Becker, J. Polchinski and A.A. Tseytlin,
 Phys. Rev. D56 (1997) 3174, hep-th/9706072. 
\bibitem{pol1} J. Polchinski and P. Pouliot, Phys. Rev. D56 (1997) 6601,
 hep-th/9704029. 
\bibitem{matt} N. Dorey, V.V. Khoze and M.P. Mattis, 
 Nucl. Phys. B502 (1997) 94, hep-th/9704197.
\bibitem{suu} T. Banks, W. Fischler, N. Seiberg and L. Susskind,
  Phys. Lett. B408 (1997) 111, hep-th/9705190. 
\bibitem{40} I. Chepelev and A.A. Tseytlin, hep-th/9709087.
\bibitem{41} E. Keski-Vakkuri and P. Kraus, hep-th/9709122.
\bibitem{gib} G.W. Gibbons and D.L. Wiltshire, Ann. Phys. 167
  (1987)  201, Erratum Ann. Phys. 176 (1987) 393.
\bibitem{rus} J.G. Russo and A.A. Tseytlin, 
  Nucl. Phys. B490 (1997) 121, hep-th/9611047.
\bibitem{duff} M.J. Duff and K.S. Stelle, \pl{B253}{91}{113}.
\bibitem{tow} P.K. Townsend, \pl{B373}{88}{545}.
\bibitem{lu} M.J. Duff and J.X. Lu, \np{B390}{93}{276}.
\bibitem{yyy} C. Schmidhuber, \np{B467}{96}{146}. 
\bibitem{st} G. Horowitz and A. Strominger, \np{B360}{91}{197}.
\bibitem{she} H. J. Sheinblatt, hep-th/9705054.
\bibitem{jjj} R. Khuri and T. Ortin, Phys. Lett. B373 (1996) 56.
\bibitem{pol} D. Pollard, J. Phys. A16 (1983) 565.
\bibitem{gro} D.J. Gross and M.J. Perry, Nucl. Phys. B226 (1983) 29.
\bibitem{sor} R. Sorkin, Phys. Rev. Lett. 51 (1983) 87.
\bibitem{mall} J.M. Maldacena, hep-th/9607235.
\bibitem{tay} W. Taylor, Nucl. Phys. B508 (1997) 122, hep-th/9705116.
\bibitem{mal3} J.M. Maldacena, hep-th/9709099.
\bibitem{egkr} S. Elizur, A. Giveon, D. Kutasov amd E. Rabinovici,
  Nucl. Phys. B509 (1998) 122, hep-th/9707217.
\bibitem{lms} A. Losev, G. Moore and S. L. Shatashvili,
  hep-th/9707250.
\bibitem{bk} I. Brunner and A. Karch, hep-th/9707059.
\bibitem{hg} A. Hanany and G. Lifschytz, hep-th/9708037.
\bibitem{brs} M. Berkooz, M. Rozali and N. Seiberg, 
 Phys. Lett. B408 (1997) 105, hep-th/9704089.
\bibitem{sei} N. Seiberg, Phys. Lett. B408 (1997) 98, hep-th/9705221.
\bibitem{sen} A. Sen, hep-th/9709220.
\bibitem{sei2} N. Seiberg, Phys. Rev.Lett. 79 (1997) 3577, hep-th/9710009.

\end{thebibliography}
\end{document}